# A Software Package for Rigorously Calculating Optical Plasma Spectra and Automatically Retrieving Plasma Properties

Xiaofeng Tan*

X Scientific, Inc., 1 Bramble Way, Acton, MA, 01720, USA.

**ABSTRACT:** In this article, a software package code named OPSIAL (Optical Plasma Spectral Calculation And Parameters Retrieval) for rigorously calculating optical plasma spectra and for automatically retrieving plasma parameters is presented. OPSIAL calculates the absolute spectral radiance caused by the bound-bound transitions of elemental species in the plasma by rigorously solving the equation of radiative transfer using an ultrafast line-by-line algorithm. OPSIAL supports both the local-thermodynamic-equilibrium (LTE) or partial LTE conditions and takes account of line broadenings due to the Doppler effect and collisions with electrons and other pseudo colliders in the plasma. An algorithm for fully automatically identifying elemental species and retrieving plasma parameters based on observed plasma emission spectra has been implemented into OPSIAL. The structure and theoretical framework of OPSIAL, together with a case study of using OPSIAL to analyze laser-induced breakdown spectral data of the ChemCam instrument onboard the Mars rover Curiosity, are presented.

## 1. Introduction

Accurate calculation of optical emission spectra of plasma is one of the key steps for understanding the plasma physics and chemistry and has tremendous applications in both fundamental research fields such as astrophysics, combustion, propulsion, plasma dynamics and in numerous practical spectroscopic diagnostic and analysis techniques such as optical emission spectrometry,[1] inductively coupled plasma,[2] and laser-induced breakdown spectroscopy (LIBS).[3] Rigorous treatment of the plasma optical emission problem requires the solution to the equation of radiative transfer[4] along the line-of-sight (LOS) through the plasma. In the general case where there are a significant number of emission lines that are not optically thin, the solution to the equation of RT demands significantly amount of computational power. For this reason, most software tools for treating plasma optical emission take the shortcut by assuming the optical thin condition that leads to significant reduction to the computational power at the cost of losing accuracy and generality for treating the emission problem.

In a recent paper, Tan introduced an ultrafast line-by-line algorithm for solving the equation of radiative transfer for gaseous emitters with large number of emission lines.[5] The algorithm is based on calculation of the optical depth in the Fourier space that reduces the time-consuming integration of the optical depth along the LOS to multiplications and therefore speeds up the calculation by two orders of magnitude or more depending on the number of lines involved. More speedup can be achieved if this method is combined with a binning technique as demonstrated by Tan. The major motivation for the development of the OPSIAL package is to apply this new development to the plasma emission problem to achieve the goal of speed, accuracy, and flexibility in the treatment of the problem. To achieve this goal, OPSIAL also rigorously treats line broadenings caused by the Doppler effect and collisions with the electrons and other pseudo colliders in the plasma and supports spectral calculations under the partial LTE conditions where the four temperatures of the plasma are taken into account: gas kinetic temperature ($T_g$), electron temperature ($T_e$), ionization temperature ($T_i$), and excitation temperature ($T_{exc}$).

Taking advantage of its ultrafast speed and high accuracy for rigorously calculating plasma optical emission, another goal attempted in OPSIAL is to achieve fully automatic determination of elemental species and retrieval of plasma parameters from plasma optical emission measurements. To the best of our knowledge, there is no method up to date that can be used to reliably accomplish this challenging task due to the complexity nature of the problem. Traditionally, species identification and retrieval of plasma parameters were performed in a series of steps that eventually require some sort of human intervention.[6] Some of the parameters such as the electron tem-perature and density may be determined from information such as the line-to-continuum intensity[7-11] or the Stark line broadening.[12-17] All methods involving human intervention eventually become very inefficient and tedious as the number of lines in the emission spectrum increases and the quality of any human involved process is difficult to gauge quantitatively. In recent years, chemometric methods based on multivariate statistical analysis in various forms have been developed to partially mitigate the situation.[18-23] These statistics based methods typically require extensive training data to generate a statistical relationship that is later used to perform the species identification and/or classification task. One issue associated with these ad hoc methods is that a dedicated algorithm has to be trained for a specific application with data prepared for the application. As a result, the whole process is typically time consuming and expensive. Moreover, a species identifier based on these statistical methods is usually not as accurate as ones based on first-principles methods. Lastly, these methods normally require wavelength calibrated spectral data for correct operation and thus is very sensitive to wavelength calibration error or drift in the training data.

As a work-in-progress feature, OPSIAL addresses the automatic retrieval problem in an approach based on the construction a species identifier calibrated with a performance library (PL) that contains information on performance metrics of the species identification process. The performance metrics information in the PL is generated using training data calculated by Monte Carlo simulations with OPSIAL's fast emission calculation code. This approach enables OPSIAL to identify species in any plasma emission spectrum in a generic way provided enough training data have been generated in the simulations. Once the species are identified, a built-in spectral fitter in OPSIAL can be used to automatically determine plasma parameters by minimizing the difference between the fitted and the observed plasma emission spectra.

In this paper, we present the structure, theoretical basis, and workflows of the OPSIAL package. The utility of the OPSIAL package is demonstrated in a case study where OPSIAL is used to analyze the LIBS spectral data taken by the ChemCam instrument onboard Mars rover Curiosity.[24] Free version of OPSIAL can be obtained from http://www.cyber-wit.com/opsial.

## 2. Structure, Theory, and Performance

### 2.1 High-level Program Structure

OPSIAL is comprised of two major subsystems: the Spectral Calculator & Fitter (SCF) and the Automated Analyzer (AA). The SCF is the spectral calculation engine that takes the user's input of the LOS information and performs ultrafast RT calculations to obtain the absolute spectral radiance. The user has the flexibility to specify parameters for multiple LOS segments, each of which is homogeneous in terms of physical and chemical properties. The LOS parameters include the lengths of the LOS segments, plasma temperatures, pressures, chemical species and their mixing ratios. OPSIAL supports both LTE and partial LTE in which the four plasma temperatures (i.e., $T_g$, $T_e$, $T_i$, and $T_{exc}$) can be individually specified. The SCF is capable of fitting a given spectrum by optimizing the plasma parameters such as the temperature, pressure, and mixing ratios. For the purpose of simplicity and to avoid the multiple-minima issue of the local optimizer used in OPSIAL, the spectra fitter in the current version of OPSIAL (v1.1) supports only spectra fitting under the LTE condition. The AA of OPSIAL is dedicated to the automatic retrieval problem and its key component is a calibrated species identifier that automatically identifies elemental species in the plasma by matching the observed plasma emission spectrum with extracted spectral features saved in a Reference Spectral Library (RSL). Once the elemental species in the plasma are successfully identified, the AA invokes the SCF to automatically optimizing plasma parameters by minimizing the difference between the fitted and the observed spectra.

All the calculation code of OPSIAL is coded in the C++ programming language for fast performance. The OpenMP (Open Multi-Processing) technology is employed in OPSIAL to parallelize time-consuming spectral calculation code. While running on a shared-memory multi-core system, OPSIAL automatically detects the number of cores in the system and creates appropriate number of threads for the spectral calculation logic and distributes the calculation tasks evenly among the cores. OPSIAL also features a user-friendly graphical front end (GFE) that is developed in the Java[TM] programing language. All user input to the program are received through the GFE. The GFE features a number of tabbed windows for outputting run-time messages, calculation results, and plots of the input and the calculated spectra. A screen shot of OPSIAL is shown in Fig. 1.

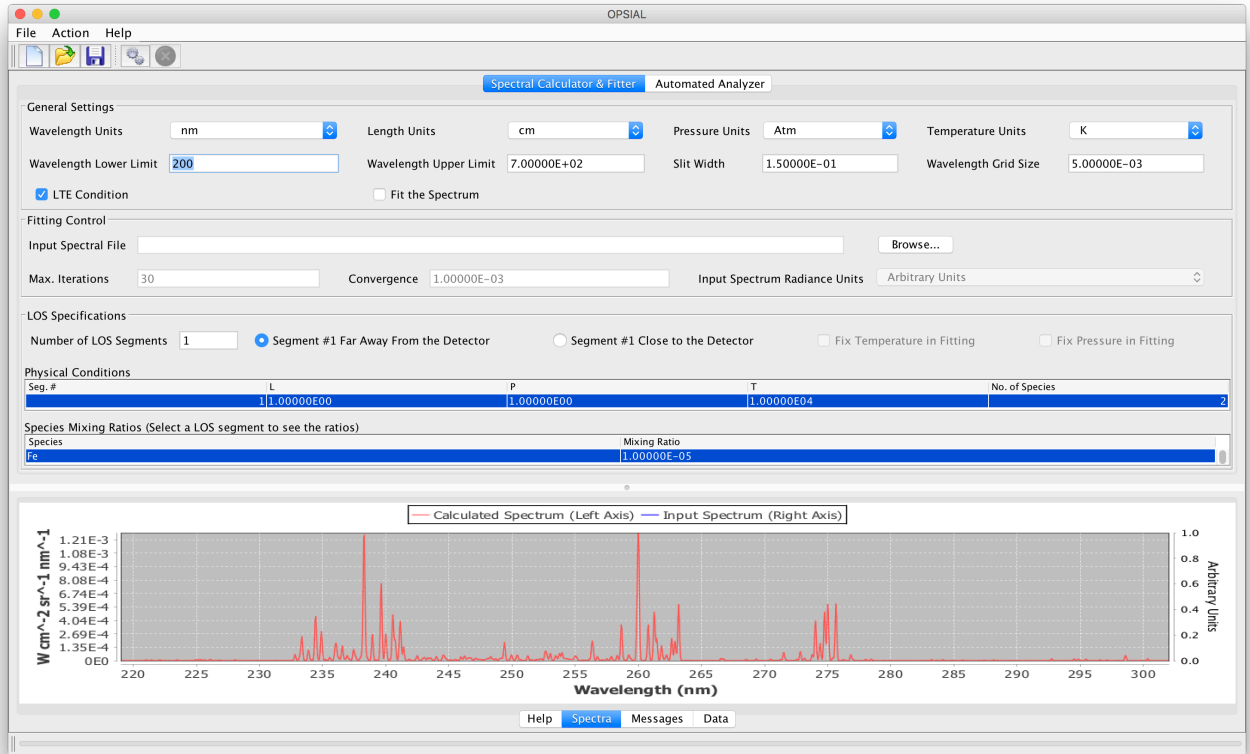

Figure 1. A screenshot of the OPSIAL program.

2.2 RT Algorithm

Tan's ultra-fast line-by-line radiative transfer algorithm is used to quickly and accurately calculate emission spectra of the plasma.[5] The algorithm is based on fast convolution of spectral lineshape using the fast Fourier transform (FFT) method and is described briefly as follows. Solutions to the equation of radiative transfer in the plasmas requires calculation of optical depth as a function of the path though the LOS, which in numeric calculations is divided into a number of homogenous segments with identical physical and chemical properties. In a line-by-line algorithm, the wavelength-dependent optical depth caused by the bound-bound emission of elemental species for a homogeneous LOS segment is given by:

$$\tau(\tilde{v}; T, P, U_1, U_2, \cdots, U_{N_s}) = \sum_{\alpha}^{N_s} U_\alpha \sum_i^{N_\alpha} S_{\alpha,i} V(\sigma_{\alpha,i}, \gamma_{\alpha,i}; \tilde{v} - \tilde{v}_{\alpha,i}), \qquad (1)$$

where $\tau(\tilde{v}; T, P, U_1, U_2, \cdots, U_{N_s})$ is the optical depth that is a function of the wavenumber $\tilde{v}$, plasma temperature $T$, total pressure $P$, and the column densities $U_1, U_2, \cdots, U_{N_s}$ of the $N_s$ elemental species in the segment; $S_{\alpha,i}$ is the line strength for the $i$-th line of species $\alpha$ that has $N_\alpha$ lines in total; $V(\sigma_{\alpha,i}, \gamma_{\alpha,i}; \tilde{v} - \tilde{v}_{\alpha,i})$ is the lineshape function for the $i$-th line of species $\alpha$ whose central wavenumber is $\tilde{v}_{\alpha,i}$; $\sigma_{\alpha,i}$ and $\gamma_{\alpha,i}$ are the Doppler width and the collision broadened half width (HWHM) for the $i$-th line of species $\alpha$, respectively. In OPSIAL, the elemental species in different ionization states are assumed to be in the Saha ionization equilibrium and a built-in ionization solver is used to solve for the ionization ratios for all elemental species.

The line strength $S_{\alpha,i}$ in eq. (1) is given by:

$$S_{\alpha,i} = S_{\alpha,i}^0 \frac{Q_\alpha^0}{Q_\alpha(T)} exp\left[-c_2 E_{\alpha,i}\left(\frac{1}{T} - \frac{1}{T^0}\right)\right]\left[\frac{1-\exp\left(-c_2 \tilde{v}_{\alpha,i}/T\right)}{1-\exp\left(-c_2 \tilde{v}_{\alpha,i}/T^0\right)}\right], \qquad (2)$$

where $S^0_{\alpha,i}$ is the line strength at a reference temperature $T^0$ (i.e., 296 K); $Q^0_\alpha$ and $Q_\alpha(T)$ are the total partition functions of the gas species at the reference temperature and the temperature of $T$, respectively; $E_{\alpha,i}$ is the total lower state energy of the spectral line; $c_2$ is a constant that equals to $hc/k_B$, where $h$ is the Planck constant and $c$ is the speed of light in the vacuum and $k_B$ is the Boltzmann constant. OPSIAL incorporates all atomic transitions found in the NIST Atomic Spectra Database[25] and the Atomic Spectral Line Database of Kurucz in the spectral wavelength range between 150 nm – 1000 nm.

For the purpose of calculating plasma optical emission, it is assumed in OPSIAL that the Doppler and collision broadening are the only line broadening mechanisms and as a result the lineshape in the emission spectrum is approximated by the Voigt function. The Dopler width $\sigma_{\alpha,i}$ is given by:

$$\sigma_{\alpha,i} = \sqrt{\frac{k_B T}{m_\alpha}} \frac{\tilde{v}_{\alpha,i}}{c}, \tag{3}$$

where $m_\alpha$ is the mass of species $\alpha$. The collision broadening of spectral lines are attributed to the collisions of the emitter species with all species (i.e., pseudo colliders) and electrons in the plasma. The broadening of the emitter species with the pseudo colliders is approximated as:[5]

$$\gamma_{\alpha,i} = 0.08 \left(\frac{T^0}{T}\right)^{0.7} P, \tag{4}$$

where $T$ is the temperature of the plasma in K and $P$ is the pressure in atm. In eq. (4), it is assumed that all emitter species in the plasma are subject to a same broadening with the pseudo colliders in the plasma. The collision broadening of the emitter species with electrons is obtained from theoretical values in the Stark-B database,[26-32] in which the broadening is fitted to an empirical function of the plasma temperature and the electron density.

Instead of calculating the optical depth directly using eq. (1), which requires a lot of computational power when $N_\alpha$s are large, the solution to eq. (1) in the Fourier space is sought in Tan's algorithm. Eq. (1) can be easily rewritten in the Fourier space using the convolution theorem:

$$\tilde{\tau}(k; T, P, U_1, U_2, \cdots, U_{N_s}) = \sum_\alpha^{N_s} U_\alpha \sum_i^{N_\alpha} S_{\alpha,i} \exp\left(-\sigma_{\alpha,i}^2 k^2 - \gamma_{\alpha,i}|k| - ik\tilde{v}_{\alpha,i}\right), \tag{5}$$

where $k$ is the Fourier transform variable of wavenumber $\tilde{v}$. If the LOS information (i.e., $T, P, U_1, U_2, \cdots, U_{N_s}$) and the spectral line information (i.e., $S_{\alpha,i}$ for all lines of all species) are known, one can sum over all the terms in the right side of eq. (5) to obtain the optical depth in the Fourier space and then performs an inverse Fourier transform to get the real optical depth.

In Tan's original algorithm, a low-resolution binning (LRB) technique is employed in addition to the Fourier transform. The principle of LRB is to bin the pressure-entangled quantities in eq. (5) so as to factor pressure-related terms out of the last summation in eq. (5) and enable a pre-computation approach that leads to further significant speedup of the algorithm. Due to the complexity of the collision broadening of the emitters with the electrons in the plasma, the current version of OPSIAL (v 1.1) does not employs the LRB technique. The speedup of the algorithm comparing to traditional line-by-line algorithms solely comes from the speedup of the calculation of the Voigt lineshape function via FFT, which already leads to more than two orders of magnitude of speedup for a typical line-by-line calculations with hundreds of lines and thousands of wavelength bins.

2.3 Retrieval of Plasma Properties

The plasma property retrieval algorithm in OPSIAL is still a work-in-progress feature and is comprised of a number of steps. The workflow of this retrieval algorithm is presented in Fig. 2, in which action objects are represented as rectangles and data objects are represented as parallelograms. For convenience, these objects are numbered and are referred to in the text with their corresponding numbers enclosed in parentheses. Important steps of the retrieval algorithm are presented in the following subsections. For simplicity the LTE condition and single LOS segment are assumed in the AA of the current version of OPSIAL (v1.1).

### 2.3.1 Spectral Feature Extraction and Matching

A spectral feature extraction (SFE, 110) method is used to extract emission features from the input spectrum. Before features extraction the SFE first performs de-noising, peak identification, and continuum baseline removal from the input emission spectrum. Peak identification is done in two steps: 1) emission peaks are identified by finding local maxima in the de-noised data; 2) the information obtained in step 1) is used by a peak fitter that fits all identified peaks into multiple Gaussian peaks. The multivariate peak fitting is intended to de-convolute overlapped peaks so as to accurately determine the positions, heights, and widths of each identified peaks. The positions, heights, and widths of the identified peaks form the extracted spectral features (120).

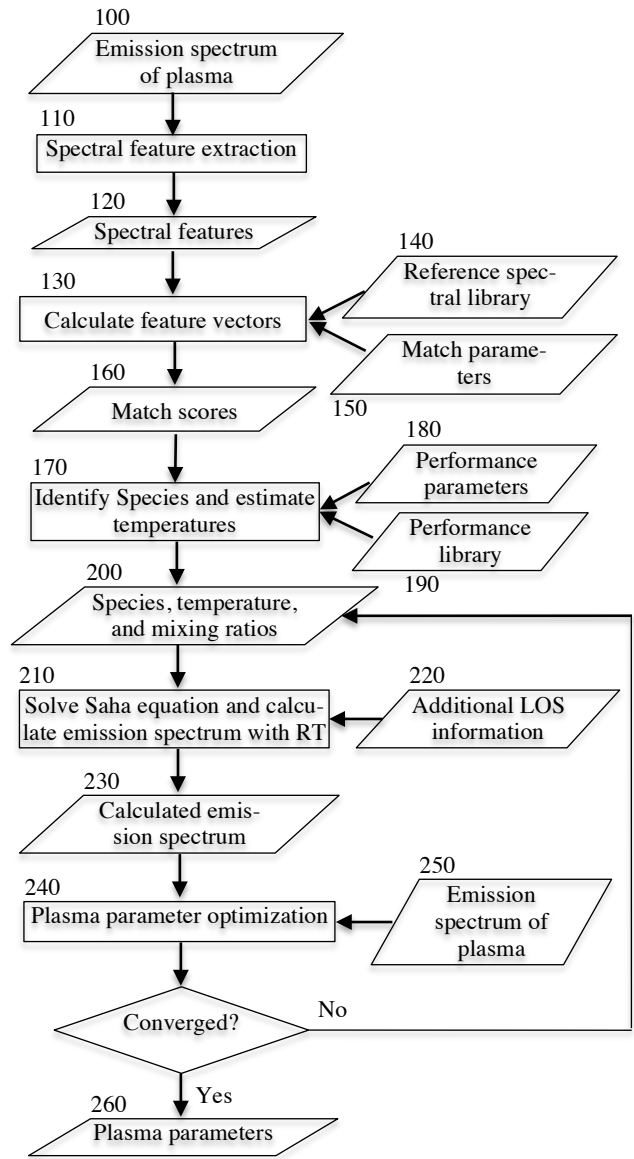

Figure 2. Workflow of the plasma property retrieval algorithm of OPSIAL. Action objects are represented as rectangles and data objects are represented as parallelograms.

OPSIAL uses the built-in RSL (140) to facilitate spectral matching. The RSL is comprised of a collection of key and value entries for all supported elemental species, with each entry corresponds to an elemental species at a specific plasma temperature. For a specific elemental species, the key field in the RSL is a tuple of plasma temperature and instrument slit width while the corresponding value field is a list of spectral features (120) extracted from the calculated spectrum using the corresponding key field as the input to OPSIAL together with other LOS parameters such as LOS length and pressure. In the current version of OPSIAL (v1.1), the temperature range used in the RSL is from 800 K to 30000 K with a step size of 1000 K while the slit width ranges from 0.008 $cm^{-1}$ to 2.6

cm$^{-1}$ in a geometric sequence with a common ratio of 2. For simplicity, the slit function is assumed to be Gaussian. The default values for LOS length and pressure are set to 1 cm and 1 atm, respectively.

The extracted spectral features (120) of the input spectrum are compared with the spectral features in the entries of the RSL. Two peaks are considered a potential match if the difference of the peak positions is within a specific wavelength threshold, or the wavelength match threshold (WMT). An elemental species is considered a candidate species in the plasma if a specific number of peaks, or the peak number threshold (PNT), of the species are found to be potential matches to some entries in the RSL. For each candidate species, the linear correlation coefficient of the matched peak positions is further checked against a specific threshold, or the correlation threshold (CT). The candidacy of the species is revoked if this check fails. For a qualified species, two feature vectors (130) are formed: one is comprised of the peak intensities in the RSL and the other of the peak intensities extracted from the input spectrum, respectively. A match score (160) is calculated for the two feature vectors using the following equation:

$$S = \left(1 - \frac{N}{N_1 + N_2}\right) \cdot \frac{\sum_{k=1}^{N} X_k^{(1)} X_k^{(2)}}{\sqrt{\sum_{k=1}^{N} X_k^{(1)^2}} \sqrt{\sum_{k=1}^{N} X_k^{(2)^2}}}, \quad (6)$$

where $X_k^{(1)}$ and $X_k^{(2)}$ are the integrated intensities of the peaks in the input spectrum and in the RSL entry, respectively; $N_1$ and $N_2$ are the number of peaks in the input spectrum and in the RSL entry, respectively; and $N$ is the number of unique peaks present in both the input spectrum and the RSL entry. It can be seen that eq. (6) is the cosine similarity between the two feature vectors times the ratio of the number of common peaks over the total number of peaks.

2.3.2 Performance Library

The PL is used to facilitate species identification and is constructed through Monte Carlo simulations. In each of such simulations, one or multiple elemental species are randomly selected with randomly generated mixing ratios and plasma temperatures. The results of the simulations are used to calculate two performance metrics for the species identifier: the True Positive Rate (TPR) and the False Positive Rate (FPR). TPRs and FPRs are functions of the match score (160) and are calculated through the receiver operating characteristic curve.[33] The obtained TPRs and FPRs are fitted to empirical expressions for each elemental species as follows:

$$TPR = (1 - S) \cdot e^{-S/\tau_{TPR}}, \quad (7)$$

and

$$FPR = (1 - S) \cdot e^{-S/\tau_{FPR}}, \quad (8)$$

where $S$ is the match score; $\tau_{TPR}$ and $\tau_{FPR}$ are constants to be derived in the fitting and saved in the PL. As an example, the calculated and fitted TPRs of Ca as a function of the match score is shown in Fig. 3.

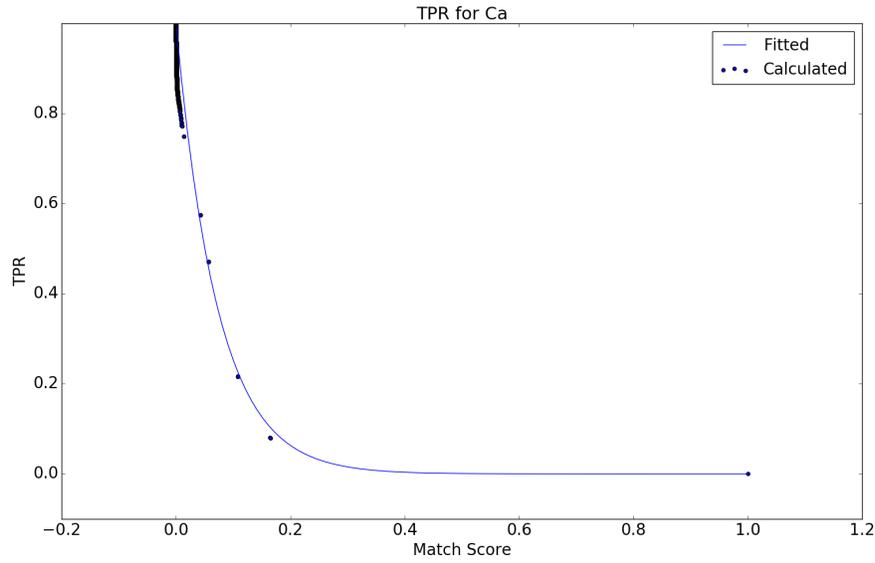

Figure 3. Fitting of the TPR as function of the match score for Ca.

For each species, a cutoff match score $S_{cutoff}$ is determined according to the following equation:

$$TPR + FPR = (1-S)\left(e^{-S_{cutoff}/\tau_{TPR}} + e^{-S_{cutoff}/\tau_{FPR}}\right) = R, \qquad (9)$$

where $R$ is a cutoff threshold specified by the user. Once the cutoff match score $S_{cutoff}$ is determined for each elemental species, the match score is converted into match index according to the following equation:

$$I = S/S_{cutoff}. \qquad (10)$$

A schematic of the distributions of the match score for the condition negative (i.e., the species is not present in the plasma) and condition positive (i.e., the species is present in the plasma) cases is shown in Fig. 4. It is shown in Fig. 4 that a larger cutoff threshold $R$ in eq. (9) pushes the line of $S_{cutoff}$ to lower values and vice versa.

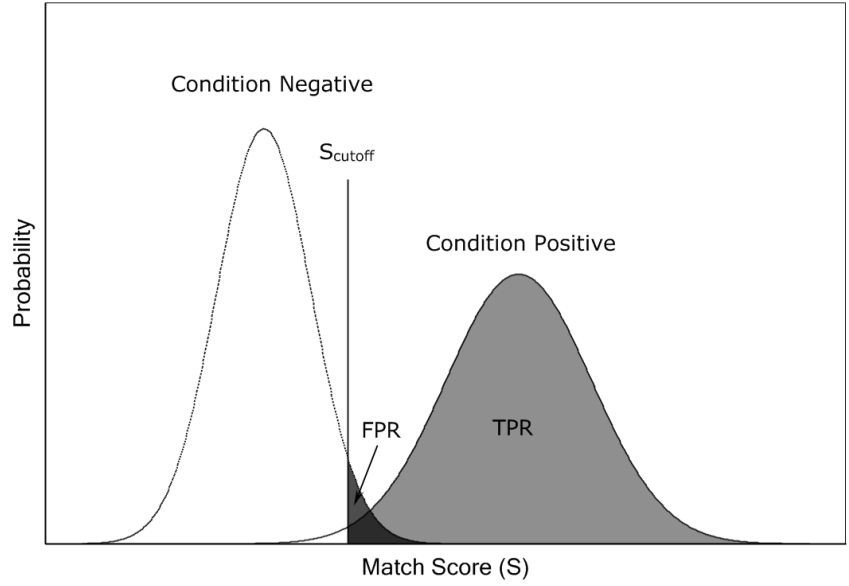

Figure 4. Schematic of match score distributions for the condition negative and condition positive cases. TPR and FPR are the light and dark shaded areas, respectively. The sum of the two is the cutoff threshold $R$ in eq. (9) that determines the cutoff match score $S_{cutoff}$.

2.3.3 Species Identification and Parameter Retrieval

To identify elemental species in the plasma, OPSIAL calculates the overall match score (OMS) for all the species that pass the cutoff match scores as determined in eq. (9):

$$OMS(T; \Delta w) = \sum_\alpha S(\alpha; T; \Delta w) / N_\alpha, \tag{11}$$

where $S(\alpha; T; \Delta w)$ is the match score for candidate species $\alpha$ for plasma temperature $T$ and wavelength shift $\Delta w$; $N_\alpha$ is the total number of matched RSL entries in the summation in eq. (11) for species α. For a specific species, there may be multiple matched entries in the RSL corresponding to multiple temperatures satisfying the cutoff match score. During the species identification, OPSIAL shifts the extracted peak positions by adding to it a wavelength shift $\Delta w$ in a user specified wavelength range to see whether higher *OMS'* can be produced. The reason for doing this is to enable OPSIAL to automatically correct for wavelength error in the input spectrum. The optimized wavelength shift $\Delta w_0$ is obtained by finding the $\Delta w$ value that produces the largest *OMS* in eq. (11). After $\Delta w_0$ is determined, OPSIAL estimates the plasma temperature using a match-score weighted average of all the temperatures in the *OMS'*:

$$T_o = e^{\sum_T OMS(T; \Delta w_o) \cdot \log(T) / \sum_T OMS(T; \Delta w_o)}, \tag{12}$$

where $T_o$ is the estimated plasma temperature.

Once the species are identified and the wavelength correction is estimated, the estimated temperature $T_o$, together with initial mixing ratios (200) and the default LOS information (220) are sent to the spectra fitter of the SCF for further optimization by minimizing the difference of the calculated and the input spectra through a gradient-descent based optimizer (240). The final mixing ratios and plasma temperature are determined when the optimization converges. The electron density of the plasma is also obtained by solving the Saha ionization equation under these conditions.

2.4 Applications and Performance

Applications, potential usage scenarios, and performance of OPSIAL are demonstrated in this section with a test analysis of the LIBS data taken by the ChemCam instrument onboard the Mars Science Laboratory (MSL) on the rover Curiosity during a campaign at Gale crater. The specific data record used in this analysis is cl5_436550112ccs_f0211572ccam01440p3. The spectrum is the mean of 30 spectra acquired with successive co-aligned laser shots on the target.[24] Spectral data in the record taken with the UV spectrometer covering the 240 – 340 nm spectral range contain most of the emission features of the entire spectrum of the target and are used in this analysis. The measured spectrum was wavelength calibrated and the intensity was photo-metrically calibrated into pseudo spectral radiance units (i.e., in units photon $mm^{-2}$ $sr^{-1}$ $nm^{-1}$). We converted the intensity into the standard spectral radiance units used in OPSIAL (i.e., W $mm^{-2}$ $sr^{-1}$ $nm^{-1}$) with the information that the signal integration time is 3 ms/shot in ChemCam. Given the fact that the integration time is much longer than the normal lifetime of the evolving plasma the obtained spectrum is actually a temporal average of the whole LIBS emission process. The converted ChemCam spectrum is shown in Fig. 5 as the red curve.

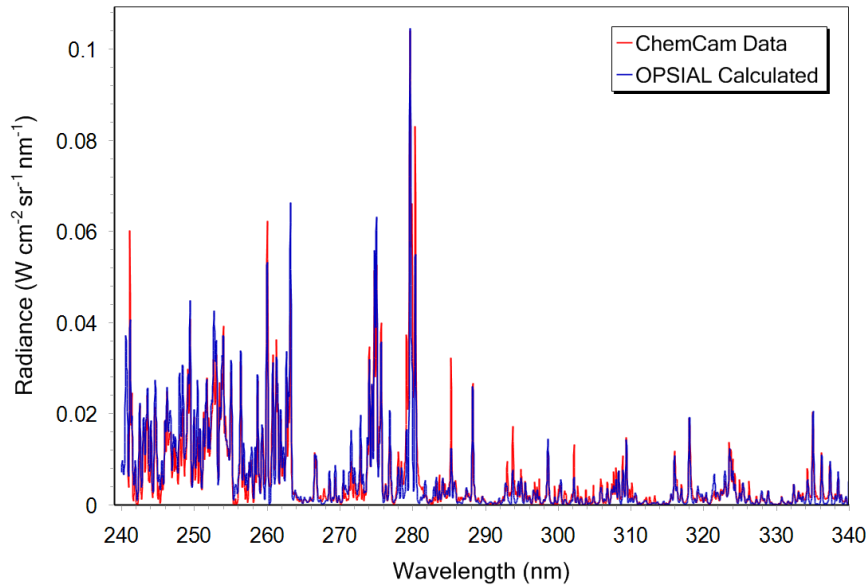

Figure 5. A LIBS emission spectrum (the red curve) taken by ChemCam during its campaign at Gale crater on Mars and the OPSIAL simulated spectrum (the blue curve).

2.4.1 Spectral Calculation and Fitting

The measured spectrum in Fig. 5 was fitted with OPSIAL to determine the plasma temperatures and abundances of the elemental species in the plasma. Four Martian atmospheric species (i.e., C, O, N, and Ar) and six metal species (i.e., Fe, Si, Al, Ti, Mg, and Ca) are included in the fitting. The four Martian species do not have strong features in the concerned spectral range and are included only for the purpose for allowing OPSIAL to more accurately calculate the electron density in the plasma. Na and K also have spectral features in the concerned spectral range but are too weak to significantly contribute to the overall spectrum. As a consequence, their abundances in the plasma cannot be reliably determined and are therefore excluded in the fitting. There are over 20,000 emission lines in the OSPIAL database for the 10 elemental species in the 240 – 340 nm spectral range. It takes only a few seconds for OPSIAL to finish an iteration of rigorous calculation of spectral radiance for all the emission lines on a PC with a 2.2 GHz Intel Core i7 CPU with 4 cores. In addition to the fast calculation speed, OPSIAL's capability of producing absolute spectral radiance is another big advantage in the fitting of the photo-metrically calibrated spectrum since the absolute spectral radiance values put a very strong constraint on the plasma temperature.

It is assumed in the fitting that the plasma can be reasonably described as a single homogeneous LOS segment with a length of 3.0 mm (which is estimated using the thermal velocity of the oxygen atom at 13500 K and a plasma expansion time of ~ 1.3 μs). A slit width of 0.2 nm, as estimated from the data, is used in the fitting. The fitting was performed in two stages. In the first stage, the fitter of the SCF was used with the atmospheric species together with Fe and Ti that are responsible for most of the emission features seen in the spectrum. The relative mixing ratios of the four atmospheric species are constrained within ~ 2% of their respective atmospheric abundances. In the second stage, the rest six metal species are included and the fitting was performed manually as the error function of the optimization is not sensitive to the abundances of these six species because they contribute only a very small portion of strong emission peaks in the spectral range.

It is found that better agreement with the ChemCam spectrum, in particular for the Fe I and Fe II features, can be achieved if two temperature values, namely, 13500 K for $T_{exc}$, $T_g$, and $T_e$ and 9000 K for $T_i$, are used in the fitting. This is an indication that the plasma has not relaxed to equilibrium at the time of its peak emission power, as is the case for most LIBS plasma under short delays.[3] It is worth mentioning that variation of the electron temperature does not significantly affect the fitted spectrum due to the fact that the spectral resolution is limited by the slit width of the instrument. The fitted spectrum is plotted in Fig. 5 as the blue curve. The mixing ratios of the elemental species as determined in the fitting are shown in Table 1.

| Elemental Species | Mixing Ratio |
| --- | --- |
| C | 0.278 |
| O | 0.547 |
| N | 0.0160 |
| Ar | 0.00472 |
| Fe | 0.0709 |
| Ti | 0.00219 |
| Si | 0.0548 |
| Mg | 0.00413 |
| Al | 0.0211 |
| Ca | 0.00160 |

Table 1. Elemental mixing ratios in the plasma as determined in the fitting of the ChemCam spectrum.

2.4.2 Automatic Plasma Species Identification

The same ChemCam spectrum in Fig. 5 was also used to test the performance of OPSIAL's automatic species identification function that is the foundation of its plasma property retrieval algorithm. The following parameters were used for the species identifier in the analysis: the cutoff threshold value $R$ is set to 1.3; WMT is set to 0.01 nm; PNT is set to 2; CT is set to 0.9999; the wavelength shift $\Delta w$ range is set to ± 5 nm.

The species automatically identified by OPSIAL together with the match indices, the TPRs and FPRs calculated at the corresponding match indices are listed in Table 2. Out of the six metal species that produces significant spectral features in the measured spectrum, five are successfully identified by OPSIAL at the above identification parameters: Fe, Ti, Mg, Si, and Ca, with the match indices of 4.65, 2.09, 1.91, 1.10, and 1.09, respectively. OPSIAL also identifies O and Na that do not contribute significantly in the concerned range with match indices of 1.27 and 1.21, respectively. The only mis-identified species is He with a match index of 2.00. It should be emphasized that the TPRs and FPRs reported in Table 2 are not TPRs and FPRs calculated at the cutoff match scores but rather at the matches scores determined for the specific species in the spectrum. As a consequence, a large match index as determined for a species in the spectrum means small TPR and FPR as calculated in eq. (7) and (8).

| Identified Species | Match Index | TPR | FPR |
| --- | --- | --- | --- |
| Fe | 4.65 | 55% | 2% |
| Ti | 2.09 | 92% | 10% |
| He | 2.00 | 71% | 21% |
| Mg | 1.91 | 81% | 17% |
| O | 1.27 | 92% | 28% |
| Na | 1.21 | 89% | 32% |
| Si | 1.10 | 97% | 29% |
| Ca | 1.09 | 98% | 28% |

Table 2. Automatically identified elemental species in the ChemCam spectrum.

## 3. Conclusions

In this paper we present the structure, theoretical framework, and workflow of the OPSIAL software package for fast and rigorous calculation of plasma optical emission spectra and for fully automatic plasma properties retrieval from the observed spectra. The fast calculation speed of OPSIAL coupled with its capabilities of treating the line broadening effects, multiple LOS segments, and NLTE temperatures makes OPSIAL a powerful and flexible software tool for quickly calculating and modeling plasma optical emission spectra. The utility of OPSIAL in processing plasma optical emission data with large number of spectral features is demonstrated in the test analysis of a ChemCam LIBS spectrum taken at the Gale crater on Mars. The analysis also shows that absolute spectral radiance calculated with OPSIAL can be of great value in fitting photo-metrically calibrated spectral data.

The automatic plasma property retrieval algorithm in OPSIAL is still a work in progress. The PL of OPSIAL in the current version is currently trained with 9570 training records generated in the Monte Carlo simulations. It is demonstrated in the test analysis of the ChemCam spectrum that even with this limited number of training data OPSIAL produces encouraging results in identifying species in the test spectrum. A large training dataset together with specially designed simulations with the knowledge of the approximate compositions (e.g., what elements are likely presented) of the plasma for specific applications are expected to produce more favorable results for the species identification. The test does show potential difficulty for OPSIAL to identify species with few spectral features (e.g., Al in the test case) surrounded by large number of interference features. It is our wish to improve the performance of OPSIAL's fully automated property retrieval algorithm in our future work.

**ACKNOWLEDGMENT**

The author is very grateful to Dr. Roger Craig Wiens at the Los Alamos National Laboratory and Steve Bender at the Planetary Science Institute for the assistance with the ChemCam data.